\documentclass[manuscript,screen, authorversion]{acmart}

\setcopyright{none}
\acmConference[]
{}
{}
{}

\acmJournal{TOSEM}
\acmVolume{}
\acmNumber{}
\acmArticle{}
\acmMonth{8}
\acmYear{2026}

\makeatletter
\AtBeginDocument{%
  \fancypagestyle{standardpagestyle}{%
    \fancyhf{}%
    \renewcommand{\headrulewidth}{\z@}%
    \renewcommand{\footrulewidth}{\z@}%
    \fancyhead[LE]{\ACM@linecountL\if@ACM@printfolios\thepage\fi}%
    \fancyhead[RO]{\if@ACM@printfolios\thepage\fi}%
    \fancyhead[RE]{\@shortauthors}%
    \fancyhead[LO]{\ACM@linecountL\shorttitle}%
  }%
  \fancypagestyle{firstpagestyle}{%
    \fancyhf{}%
    \renewcommand{\headrulewidth}{\z@}%
    \renewcommand{\footrulewidth}{\z@}%
    \fancyhead[L]{\ACM@linecountL}%
    \fancyfoot[RO,LE]{\if@ACM@printfolios\small\thepage\fi}%
  }%
  \pagestyle{standardpagestyle}%
  \renewcommand\footnotetextcopyrightpermission[1]{}%
}
\makeatother

\usepackage{graphicx}
\usepackage{xcolor}
\usepackage{hyperref}

\newcounter{principleNum}
\newcommand{\principle}[1]{%
  \stepcounter{principleNum}%
  \subsection*{Principle \arabic{principleNum}. #1}%
}

\begin{document}


\title{The Software Engineering Factory: Engineering Accountability and Control in the Age of Abundant Code}
\titlenote{Submitted to the ``New Frontiers in Software Engineering'' track of ACM Transactions on Software Engineering and Methodology (TOSEM).}


\author{Ahmed E. Hassan}
\email{ahmed@cs.queensu.ca}
\affiliation{%
  \institution{Queen's University}
  \city{Kingston}
  \state{ON}
  \country{Canada}
}

\author{Zhe Chen}
\email{sirumcz@gmail.com}
\affiliation{%
  \institution{University of Victoria}
  \city{Victoria}
  \state{BC}
  \country{Canada}
}

\author{Amr E. Mohamed}
\email{amr@insighttrail.ai}
\affiliation{%
  \institution{InsightTrail Pty Ltd}
  \city{Tarrawanna}
  \state{NSW}
  \country{Australia}
}

\author{Keheliya Gallaba}
\email{gallabak@sigsoft.org}
\affiliation{%
  \institution{Queen's University}
  \city{Kingston}
  \state{ON}
  \country{Canada}
}

\author{Daniel German}
\email{dmg@uvic.ca}
\affiliation{%
  \institution{University of Victoria}
  \city{Victoria}
  \state{BC}
  \country{Canada}
}

\author{Waleed Kadous}
\email{waleedk@gmail.com}
\affiliation{%
  \institution{Cluesmith LLC}
  \city{Mountain View}
  \state{CA}
  \country{USA}
}

\author{Gustavo A. Oliva}
\email{golivax@gmail.com}
\affiliation{%
  \institution{Queen's University}
  \city{Kingston}
  \state{ON}
  \country{Canada}
}

\author{Radin Shayanfar}
\email{radin.shayanfar@gmail.com}
\affiliation{%
  \institution{Queen's University}
  \city{Kingston}
  \state{ON}
  \country{Canada}
}

\author{Paul Stack}
\email{paul@swamp-club.com}
\affiliation{%
  \institution{Elder Swamp Club Inc}
  \city{San Francisco}
  \state{CA}
  \country{USA}
}

\author{Haoxiang Zhang}
\email{haoxianghz@gmail.com}
\affiliation{%
  \institution{Queen's University}
  \city{Kingston}
  \state{ON}
  \country{Canada}
}

\renewcommand{\shortauthors}{Hassan et al.}



\begin{abstract}
AI agents can now plan, write, test, review, and revise code at a volume no human organization can inspect line by line. \textbf{Agents can produce the code, but they cannot own it.} That scale creates an ownership problem: engineering leaders and their organizations still must answer for the code’s behaviour, failures, maintenance, and associated risk. The obligation has not changed, but the organizational burden of meeting it has grown beyond what final code review can carry. \textbf{Final code review therefore cannot remain the primary organizational control for owning code changes.} 

The Software Engineering (SE) Factory moves control into the process under which the code was produced. The Factory protocols define how work moves, what evidence must exist before it can proceed, and who is responsible for owning the result. A deterministic protocol engine, implemented in conventional code rather than prompts, enforces those protocols by controlling stage transitions, requiring evidence, and invoking gates and human escalation, where needed, before work can advance. The SE Factory preserves human authority over consequential decisions and makes accountable acceptance possible without requiring humans to supervise every agent action.

\textbf{The SE Factory is not a dark factory.} Humans remain in control by defining the scope and granularity of delegation, the boundaries and escalation conditions, and the consequential decisions that require accountable human judgment.

We draw on our experience building three distinct SE Factory implementations, all of which have undergone multiple iterations and two of which are publicly available as open-source systems. Although these implementations differ in design and strengths, a common set of principles has emerged across them. We distill these lessons into ten principles for conceiving, building, and operating SE Factories at scale, grounded in the organizational and technical realities of our implementations. We do not claim that these principles are exhaustive or constitute the only viable foundation for an SE Factory. Rather, we offer them as a conceptual scaffold and shared vocabulary for reasoning about SE Factories, evaluating their designs, and guiding their evolution. Our goal is to catalyze a community-wide dialogue that moves software engineering beyond controls designed for human-scale code production toward a disciplined, scalable, and trustworthy agentic future.
\end{abstract}

\maketitle

\section{Introduction}
\textbf{Final code review is dead. Accountability will never die.} AI agents can now plan, write, test, review, and revise code at a volume no human organization can inspect line by line. Yet long before AI, four letters had already undermined code review as a serious organizational control: \textbf{LGTM (Looks Good To Me)}. Modern software had already grown beyond what any code reviewer could fully understand, and every team today depends on libraries, services, and systems that no individual has ever read. Final code review can no longer serve as the primary organizational control for owning code changes.

\textbf{Review is not dead. The old final-code-review bargain is.} Classical software inspection was designed as a structured verification process with defined roles, explicit criteria, recorded defects, and follow-up~\cite{fagan1976design,gilb1993software}. The final-code-review bargain that came to dominate everyday practice was much thinner: a peer-to-peer transaction. A human reviewer helped a human teammate find defects then gave an LGTM so that teammate could carry the work forward. The author of the code remained a member of the team, with a reputation, continuing responsibilities, and an obligation to live with the consequences.

\textbf{Your agent is not your buddy; it has no skin in the game.} It cannot care about the verdict, stake its reputation on the change, answer for a failure, or accept a continuing obligation to maintain the code. Once a human accepts its output, the human and the organization inherit and own the result.

Even if an organization could scale its review capacity as quickly as agents scale code production, the old arrangement would still fail. The game has changed not only because the volume of code has exploded, but because the players and their obligations have changed. If the accepting human sees only the finished diff and cannot explain how the change was conceived, bounded, challenged, and verified, that human is being asked to own work whose production they never meaningfully controlled.

Review remains essential, but it must move inside the production process as independent challenge, validation, and accountable handover. Reading the final code may remain one valuable check, but an LGTM on the finished diff can no longer be the organizational mechanism that transforms agent output into software that an organization can responsibly own.

\textbf{Agentic Software Engineering (Agentic SE) represents a tectonic shift in how our industry produces software.} It transforms the volume, speed, complexity, and origin of code simultaneously, exposing a fundamental ownership gap: agents increasingly produce the code, but humans and their organizations remain accountable for it. That obligation has not changed; fulfilling it has become far more difficult. Earlier work framed this transition as a move toward intent-first development with AI teammates, requiring us to rethink software engineering’s foundational actors, processes, tools, and artifacts~\cite{hassan2024se3,hassan2025agentic, hassan2026agenticsebook}. Here, we examine the accountability challenge created by this transition as humans move out of the inner loop and into an emerging middle loop, where they direct, coordinate, and govern fleets of agents.

\textbf{To own a code change is to be accountable for it: to accept the risks it introduces, answer for its behaviour and failures, and take responsibility for maintaining and evolving it for as long as it remains in use.} AI agents cannot bear these obligations. Engineering leaders and their organizations must.

That ownership gap is the defining challenge facing Agentic SE and a pressing concern for engineering leaders throughout our industry. This article is written for those leaders: managers, directors, chief technology officers, and others who remain accountable for software produced by teams increasingly working with AI agents.

Abundant agentic output creates an organizational denial-of-service attack: plausible work arrives faster than the organization can absorb, scrutinize, and responsibly own. This flood overwhelms the two imperfect controls on which leaders once relied: code review and trust in people whose capability and judgment had been established over time to produce and gate the work. AI compounds the risk by making large, intricate changes cheaper to create and distribute; and Trojan horses easier to conceal within ordinary-looking code changes.

\textbf{Stop fooling yourself: you cannot code-review your way out of the AI code flood.} Engineering leaders and their organizations still must answer for the code, and the load of doing so keeps growing. Trying to solve this through more code review is like trying to make a finished skyscraper safe by checking every brick. Scaling review capacity, whether with additional reviewers or with reviewing agents, does not close the ownership gap because the gap is structural rather than a matter of throughput.

Code review cannot reconstruct missing intent, provenance, decisions, or evidence after production. Architecture decision records preserve the context and rationale for consequential design decisions~\cite{nygard2011documenting}. SLSA provenance and in-toto attestations preserve evidence about where, when, and how software artifacts were produced~\cite{slsa_framework,torres2019toto}. Source review also cannot establish trust in the entire production chain: a compromised toolchain can subvert a program without leaving evidence in the source being reviewed~\cite{thompson1984reflections}. These mechanisms embody a broader principle: evidence must be created and preserved during production because final review cannot manufacture it afterward.

Agents marking other agents' homework cannot manufacture accountability either. No reviewer agent can accept organizational risk, commit to maintaining the software, or own the result.
Software engineering is a team sport built on layers of justified trust. Yet most agentic code changes today are produced inside a private human-agent bubble: one developer and one or more agents jointly handle intent, design, implementation, testing, and code review. The organization sees the resulting code but not the interaction that produced it. Even the human inside the bubble rarely knows which alternatives the agents considered, which assumptions shaped the code, where the agents deviated, or what evidence supports their claims. Everyone outside the bubble sees even less.

\textbf{The answer is to stop treating final code review as the primary control for owning code changes and instead engineer the conditions under which those changes are produced.} The quality of the output must not depend on trusting the person or the agent. \textbf{Instead, it must depend on how we engineer the workflow that produces it.}

The Software Engineering (SE) Factory operationalizes this shift by moving control into the process under which the code is produced. Factory protocols define how work moves, what evidence must exist before it can proceed, and who is responsible for owning the result. A deterministic protocol engine, implemented in conventional code rather than prompts, enforces those protocols by controlling stage transitions, requiring evidence, and invoking gates and human escalation where needed before work can advance. The result is not merely coordinated agent activity, but visible and governed production that makes accountable acceptance possible.

The authors of this article have built three distinct SE Factory implementations, all of which have undergone multiple iterations and have been deployed across hundreds of teams worldwide. Two are publicly available as open-source systems: Codev, released under the Apache License 2.0, and Swamp, released under the GNU Affero General Public License v3.0 (AGPLv3): Codev: \url{https://codevos.ai/}; and Swamp: \url{https://swamp-club.com/}. Their availability allows readers to inspect, use, and evaluate concrete realizations of the ideas developed here. Each implementation has its own architecture and distinctive strengths. All, however, treat the SE Factory as an engineering system of humans, agents, and deterministic tools governed by explicit rules rather than foundation-model judgment. The Factories turn private interactions into visible, governed work. Humans retain authority over intent, consequential tradeoffs, exceptions, and associated risk, while agents work autonomously within visible, enforced boundaries.

Toyota has shown the world the value of such ``factory'' thinking~\cite{ohno1988toyota}. A factory is more than an ordered flow: documented factory lines standardize routine work, visible flow exposes defects and bottlenecks at their source, and the Andon cord lets a worker stop the line rather than pass a known problem downstream. A high-school dropout can help build an airplane we trust with our lives, not because the work is simple or the worker infallible, but because the engineering system bounds each task, supplies the correct inputs, defines and validates the handoffs between tasks, checks the results, and prevents one person's mistake from silently becoming everyone's risk. Toyota designed quality and flow to reinforce one another: abnormalities stop the line, problems are corrected at their source, and those corrections improve subsequent work. The SE Factory applies the same logic to Agentic SE.

This emphasis on learning from execution also has deep roots within software engineering. Basili's Experience Factory treats project experience, process data, failures, and lessons learned as organizational assets that should be analyzed, packaged, and reused to improve future work~\cite{basili1994experience}. The SE Factory extends this logic to the protocols through which humans and agents jointly produce software. Large-scale DevOps research similarly shows that high-performing organizations can improve delivery speed and stability together rather than treating them as an unavoidable tradeoff~\cite{forsgren2018accelerate}.

\textbf{The SE Factory is not a dark factory.} Across these implementations, the shared goal is symbiosis, not replacement. In Licklider's original formulation of human-computer symbiosis, humans set the goals, determine the criteria, and perform the evaluations, while machines perform the routinizable work that prepares the way for those judgments~\cite{licklider1960man}. The SE Factory applies this division of responsibility at the organizational scale.
Humans remain in control by defining the scope and granularity of delegation, the boundaries and escalation conditions, and the consequential decisions that require accountable human judgment. The Factory puts routine competence into the engineering system while preserving human judgment wherever accountability cannot be delegated to a machine. It reduces dependence on constant supervision and on the unrealistic belief that final code review can carry organizational control at this scale.

The SE Factory also changes how failures are handled. The Factory can reconstruct what happened, where control failed, what evidence existed, and who made the relevant decisions. That replaces retrospective blame based on overwhelming end-stage artifacts with evidence about the operation of the engineering system. Each failure can then improve the SE Factory so that the next run is less likely to fail in the same way.
This practitioner-led report from the field is written for researchers and practitioners alike. Its argument rests on three forms of support: two quantitative results showing that, in each case, work produced by the same underlying foundation model became substantially more reliable when governed by an engineered Factory protocol enforced by a deterministic protocol engine; the operational experience of building and iterating three distinct SE Factory implementations deployed across hundreds of teams worldwide; and convergence with five decades of software engineering precedent, which we trace throughout the ten principles.

We have made a deliberate choice in presenting this work. Alongside conventional argument and evidence, this article uses a direct voice, visual metaphors, and conceptual illustrations to make its ideas accessible across software engineering research and practice. The gap between these communities has widened at precisely the moment when Agentic SE requires them to work together most closely. Research and practice are two sides of the same discipline, and neither can solve the accountability problem in the agentic SE era alone. We therefore present these principles in a form intended to support critical examination, practical adaptation, and further development by both communities.
We do not claim that these ten principles are the final word. Others will emerge as more Factories are built, used, and revised. This article marks a point of convergence: the principles these three implementations share and that we believe are essential for SE Factories to succeed at scale. Our aim is not to prescribe a single Factory architecture, but to establish a common understanding and shared language that can support a structured and productive industry-wide discussion about Software Engineering Factories. More fundamentally, we argue that factory thinking must become a core part of how the software industry reasons about, governs, and takes responsibility for software production in an era of abundant code.

\begin{figure}[ht]
\centering
\includegraphics[width=\linewidth]{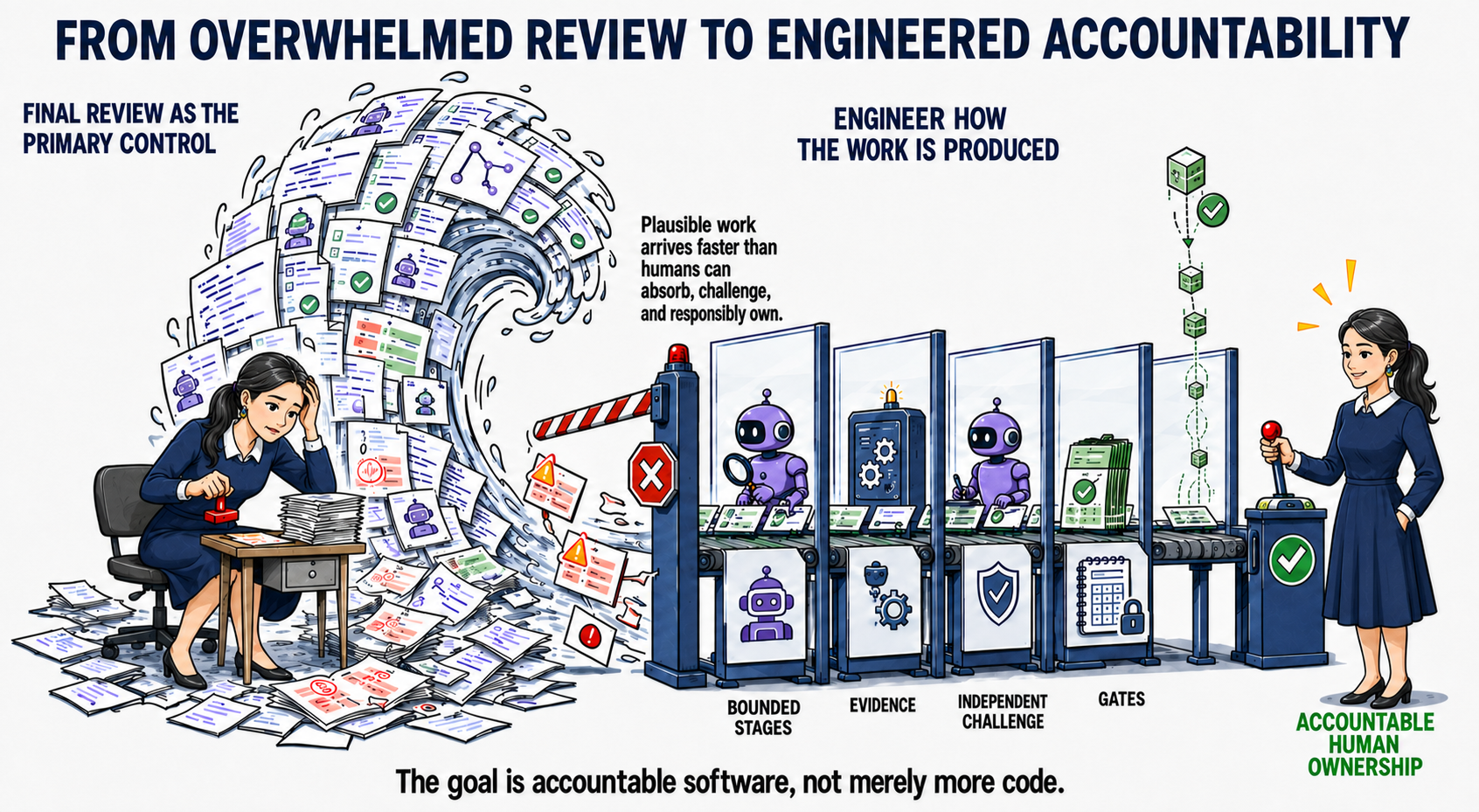}
\caption{From final-diff review to engineered accountability through bounded work, validated handoffs, evidence, challenge, gates, escalation, and accountable acceptance.}
\Description{}
\label{fig:overview}
\end{figure}
\section{Coordination Is Not Control: A Workflow Is Not a Protocol, and a Graph Is Not a Factory}

Orchestration coordinates execution; it does not establish engineering control. Harnesses, loops, graphs, and workflows provide increasingly sophisticated ways to sequence and route agentic work. Multi-agent frameworks such as MetaGPT~\cite{hong2024metagpt}, ChatDev~\cite{qian2024chatdev}, and AutoGen~\cite{wu2023autogen} assign agents software-engineering roles and route work between them, and agent-computer interfaces such as SWE-agent~\cite{yang2024sweagent} give a single agent effective access to repositories and tools. These systems help orchestrate work between agents; however, they do not govern whether work may advance, what must be established before it does, and who is authorized and accountable for accepting the result. Protocols provide that structure, and the SE Factory integrates them into a unified engineering system.

To provide that control, a Factory protocol defines the responsibility of every stage of a protocol, the inputs it may consume, the outputs it must produce, the evidence required for its claims, the gates that control advancement, the considerations that must be surfaced, the conditions that require escalation to a higher entity (e.g., a human or a more capable agent), and the authorized entity which decides what happens next. A graph or a workflow can coordinate work without establishing any of those responsibilities. It can show where work goes without establishing why it deserves to advance or who can accept the result.

A deterministic protocol engine enforces these requirements. It is conventional software, not an agent. It moves work through the prescribed stages, requires evidence at each stage, and invokes the relevant gates. Work advances only when those gates authorize it. Agents cannot rewrite the protocol, skip a gate, or move work to the next stage on their own.

A protocol may contain or invoke sub-protocols, allowing larger factory lines to compose reusable controlled protocols without losing their individual responsibilities, evidence requirements, gates, or accountability.

The term protocol comes deliberately from the medical field, where protocols combine checks that block unsafe continuation with spoken call-outs that direct expert attention while work is underway. The SE Factory makes the same distinction. A gate controls whether work may advance and may be enforced by a deterministic check, a reasoning AI-powered evaluator, or an authorized human. A steering call-out does not control advancement; it places important information or considerations into the worker's reasoning while preserving the worker's autonomy over what to do with them.

Evidence also has a precise meaning in the SE Factory. Evidence is an independently checkable support for a claim: a test result supporting the claim that tests were executed and passed, a tool record showing which files were inspected, a comparison showing whether the approved plan was followed, or an approval showing that an authorized person permitted the work to advance. An agent's statement that it completed the work is a claim, not evidence.

Much of the SE Factory depends on a strong end-to-end testing system being in place. Stage-level checks show whether individual parts behaved as expected; end-to-end tests show whether the assembled software delivers the intended outcome across its real boundaries. Without that system-level evidence, local success can conceal a failing whole. Organizations cannot introduce agents reliably without a stable outcome against which their work can be judged. Unit tests alone do not provide that boundary: they may miss user-visible behaviour and may be changed by the implementing agent until they conform to its code. The Factory therefore needs user-level, black-box, or user-acceptance tests that are not freely modifiable by that agent and serve as both an independent acceptance boundary and executable documentation of expected behaviour.

The SE Factory connects these controls through an \textbf{accountability ledger} that preserves the original intent, assignments, stage inputs and outputs, validated handoffs, evidence, deviations, interventions, escalations, gate decisions, approvals, and final accepted result. Evidence supports individual claims; the ledger preserves the chain through which the organization can reconstruct and own the outcome.

A \textbf{validated handoff} moves work from one stage to the next only after the stage output passes the checks defined by the handoff's semantic type. An \textbf{accountable handover} presents the accumulated result, and the evidence needed to judge it, to a human already authorized to decide whether the result can be responsibly owned.

\begin{figure}[ht]
  \centering
  \includegraphics[width=\linewidth]{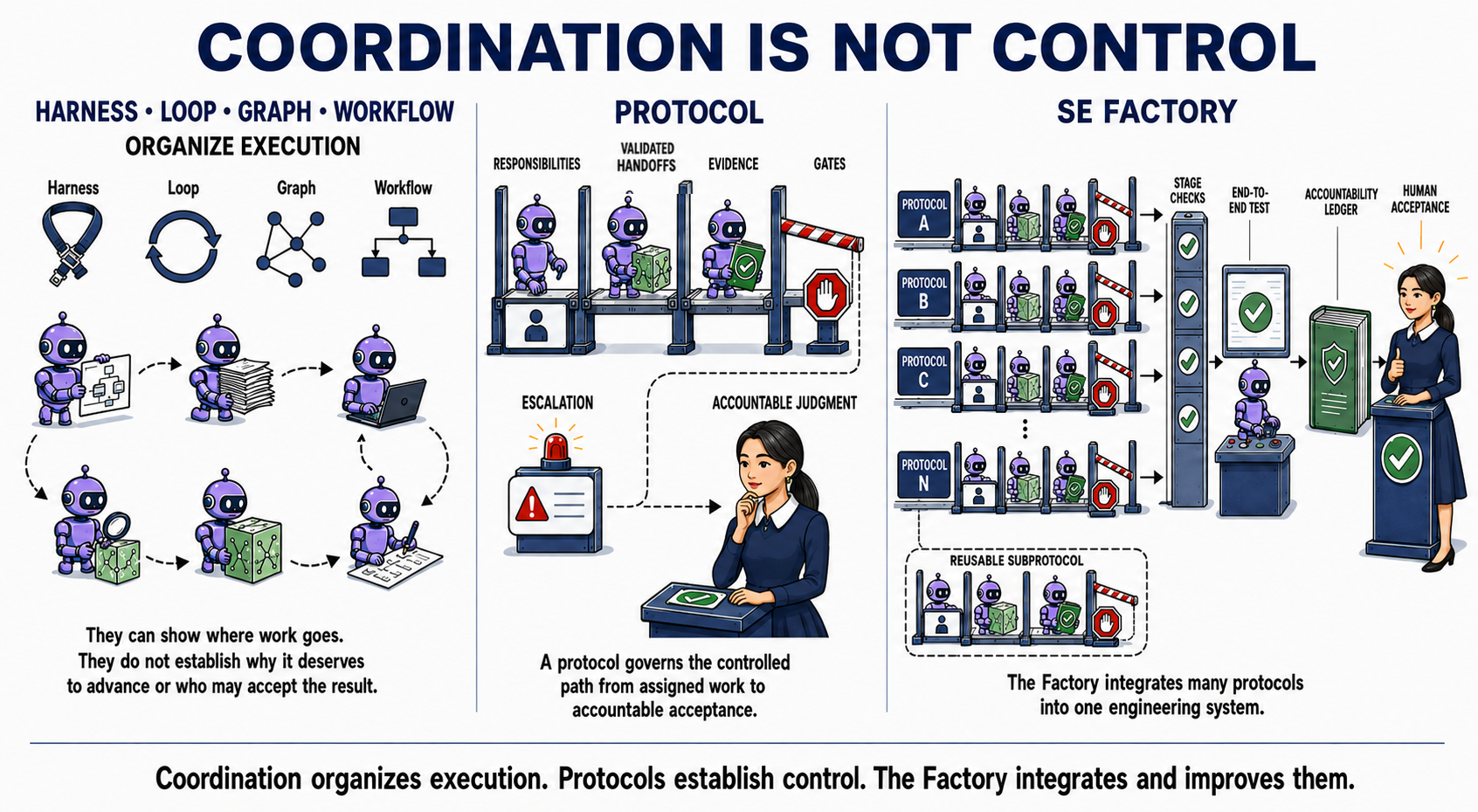}
  \caption{Coordination mechanisms organize execution; protocols govern bounded work; the SE Factory integrates, observes, and improves many protocols into one engineering system.}
  \Description{}
  \label{fig:coordination}
\end{figure}

\section{Evidence: Structure Changes What the Same Model Can Deliver}

The SE Factory claim is testable: does engineering the workflow change what the same model can deliver? Two concrete results say yes. The Alibaba group tested that proposition in code review, while one of the three Factory implementations represented here tested it in the construction of complete software projects. These two cases are the ones for which quantitative results exist; the other implementations inform the principles in this article through operational experience rather than controlled measurement.

\subsection{Alibaba Engineered Code Review as a Protocol}

Alibaba's Open Code Review, or OCR, began from a familiar industry pattern~\cite{alibaba_ocr_2024}: give a general-purpose agent a detailed skill and expect the written instructions to make the code review disciplined. Alibaba soon discovered that however emphatically instructions are phrased, a language-powered skill can only appeal for discipline; it cannot enforce it. On larger changes, the agent cut corners, reviewed only part of the work, missed relevant issues, and attached findings to the wrong locations.

Alibaba replaced their skill-powered approach with a code-review protocol. Deterministic software selects the files that must be reviewed, groups related files, matches relevant rules, and determines when a change exceeds fifty lines and therefore requires an additional risk-analysis stage. Independent AI-powered reviewing agents then examine bounded units of work with purpose-built tools and explicit budgets. A positioning stage anchors each finding to an exact location, while a reflection stage challenges weak findings and a reflection gate prevents those that do not survive the challenge from advancing. A human must approve proposed fixes, and the complete code-review session is retained.

On a benchmark of 200 real pull requests from 50 open-source repositories, labelled by more than 80 experienced engineers, OCR achieved higher precision and F1 than a general-purpose agent using the same underlying model while consuming roughly one ninth of the tokens, although it accepted lower recall as a deliberate tradeoff for fewer weak findings. Alibaba reports that the system has operated internally for two years, served tens of thousands of developers, and identified millions of defects.

The important result was not only that Alibaba OCR produced better code-review comments. The protocol made those comments more credible by governing scope, enforcing rules, grounding the analysis in explicit anchors, introducing independent challenges, and requiring human approvals. In Factory terms, code review became a governed, repeatable process, with each execution producing the evidence needed to reconstruct, audit, and improve how its conclusions were reached. This enabled organizational accountability and continuous improvement. Code review became one engineered factory line rather than an unconstrained set of interactions with one or more foundation models.

Alibaba OCR does not contradict the claim that final code review is dead as the primary organizational control. It demonstrates what review must become: a bounded, staged, evidence-producing Factory protocol rather than a final interaction in which one worker inspects another worker's finished output and supplies an LGTM.

This code-review protocol (shown below using the shared SE Factory terminology developed in this article) is only one of many protocols an organization needs. Feature implementation, bug repair, security analysis, code cleanup, and incident response also need dependable protocols, which is why the Factory must let organizations author, execute, observe, measure, and improve many factory lines.

\begin{figure}[ht]
  \centering
  \includegraphics[width=\linewidth]{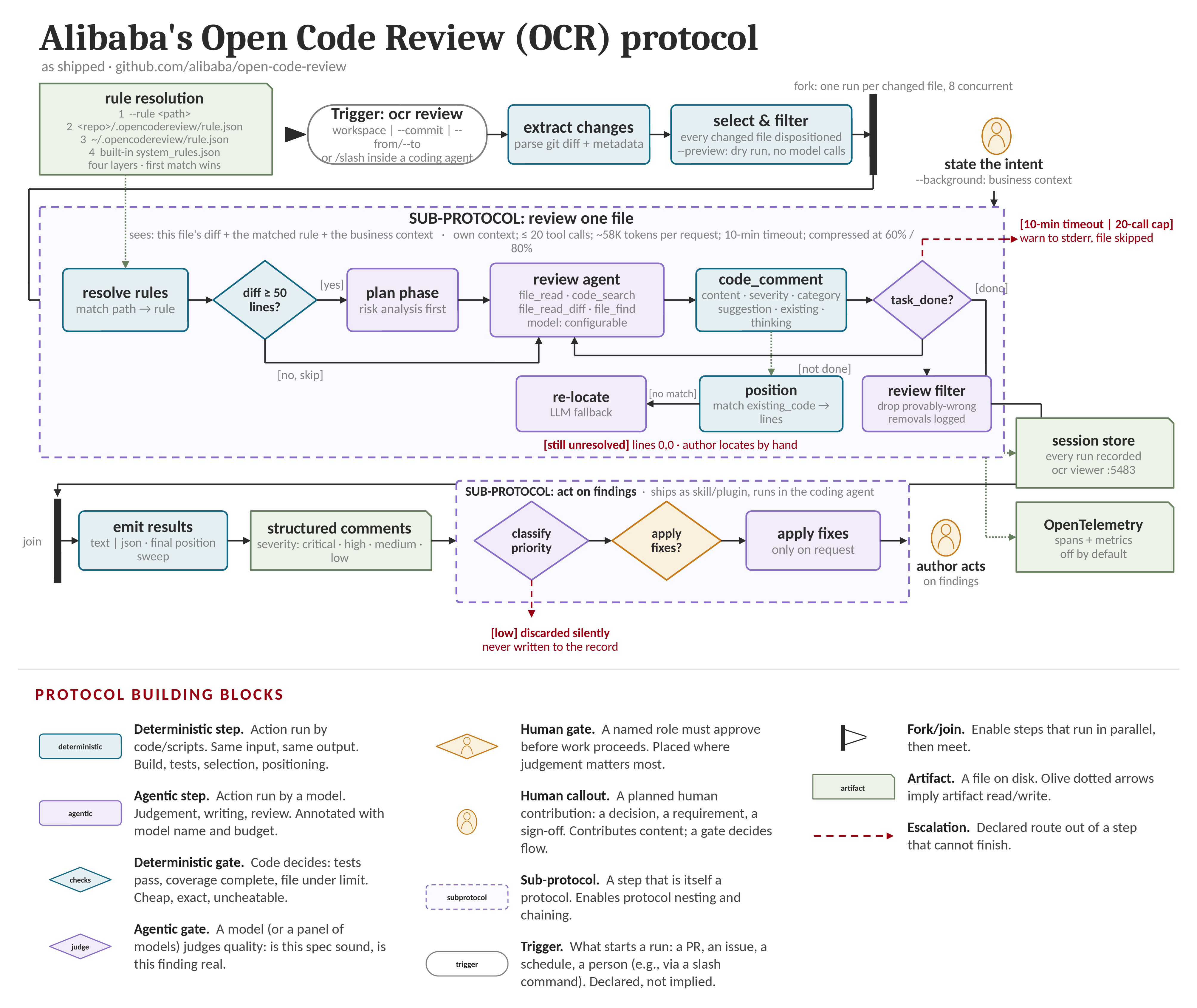}
  \caption{Alibaba's Open Code Review protocol, mapped to the SE Factory's protocol building blocks.}
  \Description{}
  \label{fig:aliyun-ocr}
\end{figure}

\subsection{The Factory Against a Single-Shot Agent}

In that Factory implementation, we tested the same proposition by asking whether a staged and gated protocol for ``specification implementation'' could outperform one capable agent using the same foundation model under the same time limit.

NL2RepoBench~\cite{ding2025nl2repo}, developed by ByteDance's Seed research team, contains 104 distinct assignments, each requiring an agent to generate a complete runnable piece of software from a written specification. The benchmark tests the construction of many complete software packages rather than repeated edits to one small program, although it does not reproduce every technical, organizational, and operational difficulty of a long-lived enterprise system.

We ran this benchmark using GLM-5 in two configurations, giving each approximately one hundred minutes per assignment. The single-shot baseline used one long conversation and was instructed to test and repair its work. The Factory used four stages: understand, skeleton, implement, and audit, with a fresh context limited to the needs of each stage. A deterministic test gate read the results from pytest, the benchmark's Python test runner, and returned failing work for correction.

The SE Factory run passed an average of 52.9 percent of tests, compared with 19.8 percent for the single-shot agent, a 167 percent relative improvement. The SE Factory run fully solved 27 assignments, compared with 8, a 238 percent relative improvement. Within the same elapsed time, the SE Factory run used approximately 2.4 times as many tokens and achieved approximately 2.7 times the test pass rate. The SE Factory exchanged additional computation for materially greater reliability rather than claiming a free efficiency gain. Only 5 of its 27 complete solutions passed with the single-shot run, so the improvement came mainly from controlled iteration rather than first-try correctness.

Both configurations encountered infrastructure errors, but the Factory persisted its state at stage boundaries and retried affected stages rather than discarding earlier completed work or restarting the entire assignment. We call this \textbf{durable execution}: completed stages remain completed, failures are localized, and interrupted work can resume from a known state. To make the comparison stricter, we removed every assignment in which the baseline run encountered any infrastructure issue; even on that cleaned-up subset, the Factory retained a substantial advantage of 57.5 percent versus 39.0 percent, a 47 percent relative improvement.

\begin{figure}[ht]
  \centering
  \includegraphics[width=\linewidth]{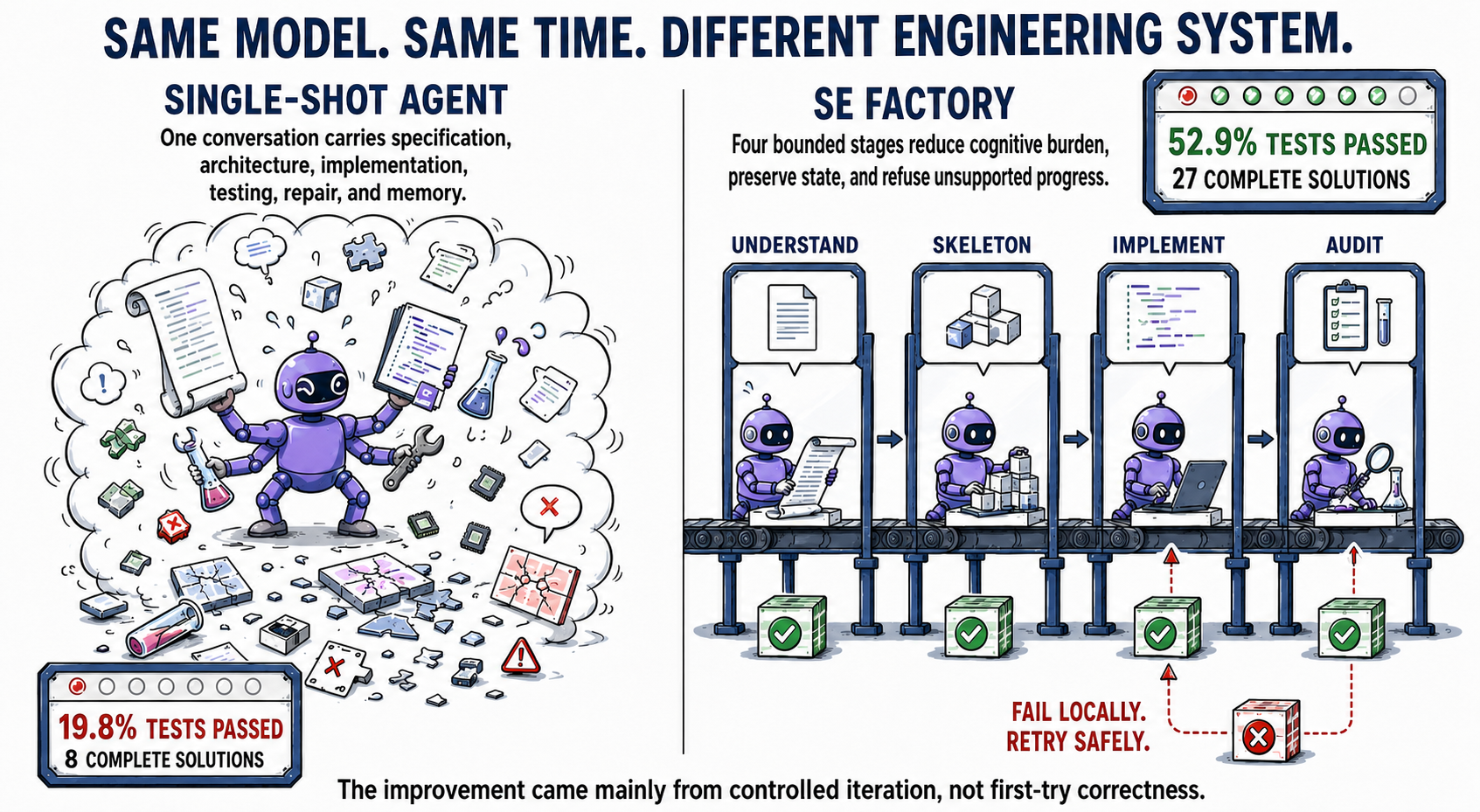}
  \caption{Same model and time limit, but a different engineering system: 19.8\% versus 52.9\% of tests passed and 8 versus 27 complete solutions.}
  \Description{}
  \label{fig:nl2repobench}
\end{figure}

This NL2RepoBench experiment is a proof of concept, not a universal law, and its limits should be stated plainly. Test pass rate is a proxy for correctness and says nothing about maintainability or architectural coherence, so the construct it measures is narrower than the quality the Factory ultimately targets. The improvement was bought with computation: the Factory consumed roughly 2.4 times the tokens of the baseline within the same elapsed time, a deliberate exchange of compute for reliability rather than a free gain. And the setting covers one benchmark, one model, and one protocol; benchmark repositories built from written specifications do not reproduce the accumulated constraints of a long-lived enterprise system, and the result does not by itself establish organizational accountability. What the experiment does show, under those stated conditions, is consequential: the same model became substantially more reliable when the engineering system reduced its burden, constrained its work, preserved its state, and refused to accept unsupported progress. The ten principles that follow explain the structure behind both results and how it extends to other Software Engineering workflows.

\section{The Ten Principles of a Factory at a Glance}

The ten principles below are the points of convergence across the three Factory implementations. They progress from the foundations of trust and ownership, through the engineering of both the work and the workforce, to the compounding long-term advantage of a Factory that improves with every run.

\noindent\textbf{THE CORE}

The first four principles establish why agentic work deserves trust, how humans and agents should divide responsibility, and who can ultimately own the result.

\begin{itemize}
    \item \textbf{Engineer Trust.} Trust should come from how work was produced and what can be proven about it, not from who or what produced it.

    \item \textbf{Face the Mess.} Design the Factory for how software is actually built: through uncertainty, iteration, failure, and change.

    \item \textbf{Cast the Minds.} Assign every participant work that fits its knowledge, strengths, limits, and authority, and size each delegation according to the risk the organization is prepared to accept.

    \item \textbf{Own What You Shape.} Do not ask a human to own an outcome unless the engineering system gives that human the needed authority, evidence, and intervention points to shape it.
\end{itemize}

\noindent\textbf{ENGINEER THE WORK}

The next three principles turn uncertain work into bounded, checkable, and economically appropriate execution.

\begin{itemize}
    \item \textbf{Stage and Type.} Divide work into stages, assign each stage a tightly scoped task, let it retrieve context on demand rather than inherit the context from prior stages, validate every handoff, and persist state so execution can retry and resume.

    \item \textbf{Gate and Steer.} Use gates to control whether work may advance and steering call-outs to place important information and considerations inside the worker's reasoning without taking away its autonomy.

    \item \textbf{Price It.} Price the workflow rather than the worker by matching protocol rigor and cost to the value of the work and the consequences of failure.
\end{itemize}

\noindent\textbf{ENGINEER THE WORKFORCE}

The next two principles organize humans, agents, and deterministic tools into a visible workforce with explicit responsibilities and bounded authority.

\begin{itemize}
    \item \textbf{Open the Floor.} Restore Software Engineering as a team sport by making live work visible so the right people and agents can join when useful or be pulled in when needed.

    \item \textbf{Anticipate, Challenge, and Contain.} Anticipate predictable human and parahuman behaviour, independently challenge consequential claims, and contain each worker's authority and failures.
\end{itemize}

\noindent\textbf{THE LONG GAME}

The final principle ensures that the value created by each execution remains in the organization and compounds into the SE Factory from one run to the next.

\begin{itemize}
    \item \textbf{Start Now and Compound.} Make every run deepen the organization’s understanding of the software system and strengthen the Factory that will perform the next run.
\end{itemize}
\section{THE CORE}

\principle{Engineer Trust}

\textbf{Trust should come from how work was produced and what can be proven about it, not from who or what produced it.}

Bridges do not stay up because of optimism, terminology, or exceptional materials. Bridges stay up because competence is built into the engineering system that produces them. The same must now become true of Agentic SE.

Building competence into the workflow raises an immediate objection: will it slow the organization down? That objection measures speed too early by counting tokens, lines, commits, or pull requests rather than dependable code delivered and safely evolved. Undisciplined speed is a loan because work skipped during design, testing, or code review returns later as failure, rework, and lost understanding. Speed is addicting precisely because its immediate reward is visible while its cost arrives later, often in another team's queue or during a production incident. This is the practical meaning of going slow to go fast: discipline at the right points reduces total delivery time and protects future velocity.

The SE Factory places competence in the protocol. A change earns trust because its intent was explicit and its scope was controlled. At every handoff, the artifact was checked against the required handoff type and prevented from advancing until it conformed. The claims of each stage were supported by evidence, the independent challenges of each stage were completed, and its gates were passed. Trust does not depend on whether a human or an agent performed the work. Software Engineering has always relied on layers of justified trust because no person can inspect or understand everything. The SE Factory rebuilds those layers around agentic output instead of leaving one developer to perform detective work at the end of a private developer-agent session that produced a mountain of code.

\noindent\textbf{Software Engineering precedent.} Watts Humphrey, the father of software process at the Software Engineering Institute, made this idea central to the Capability Maturity Model (CMM)~\cite{humphrey1989managing}: the quality of a system is governed by the quality of the software process that is used to build it. Michael Fagan's formal inspections at IBM supplied the operational proof in 1976~\cite{fagan1976design} by cutting defects through a repeatable procedure rather than through better talent. Fagan's inspections added hours early and returned them with interest by reducing rework so sharply that projects finished sooner, making quality a property of the workflow rather than the heroics of the reviewer. Barry Boehm then measured why that investment matters by showing that defects become more expensive as they survive requirements, design, implementation, testing, and release~\cite{boehm1981software}, so an issue prevented or caught at an early gate avoids a much larger downstream cost. The DevOps Research and Assessment program, known as DORA, tested the supposed tradeoff between rigor and speed across thousands of organizations~\cite{forsgren2018accelerate} and found that high-performing teams combine small batches, automated gates, continuous testing, and disciplined delivery. \textbf{Software Engineering has always been a long game: the hare may win the sprint to code, but the disciplined tortoise wins the race to trustworthy software and keeps winning as the software evolves.}

\principle{Face the Mess}

\textbf{Design the Factory for how software is actually built: through uncertainty, iteration, failure, and change.}

We keep telling ourselves a fairytale in which requirements were once complete, reviews were once thorough, and engineers once understood every line, although that past never existed. Requirements changed when users encountered working software, teams depended on components they had not read, and the living theory of a piece of software slipped out of reach decades and millions of lines ago. Software is soft: the want moves while you build, and that is precisely why making software remains difficult even after code generation has become so cheap.

Recognizing this history does not excuse low-quality agent-generated work. It changes the question from whether the worker appears trustworthy to whether the work is fit for purpose, its risks are understood, and the organization can own and maintain it. Vibe coding correctly recognizes that the mechanical cost of producing code is approaching zero. Vibe coding becomes dangerous when it abandons the harder work of determining what matters, verifying the result, and preserving a coherent software system while requirements continue to move.

The SE Factory treats discovery, failed checks, revised requirements, rejected alternatives, and departures from the approved approach as expected transitions rather than inconvenient exceptions. Each transition carries a reason and appears in the accountability ledger of the SE Factory, so the actual path remains visible rather than being replaced by an idealized narrative reconstructed after completion. The protocols of the SE Factory can loop, return, and revise without losing control because uncertainty is represented instead of hidden.

\noindent\textbf{Software Engineering precedent.} Fred Brooks split the difficulty of making software in his 1986 essay ``No Silver Bullet''~\cite{brooks1987no}. The essence is the true complexities around making software, while the accident is the mechanics of expressing it as code. Decades of languages, generators, and tools attacked the accident without removing the essence, and agentic coding is the largest reduction in accidental difficulty so far. Manny Lehman made software's softness itself a law when his research on software evolution showed that a system in active use must continually change~\cite{lehman1980programs} or become progressively less useful in its environment. Faster coding therefore cannot eliminate the essential work of deciding what to build, proving that it works, and changing it coherently as the need continues to move. Reliable Software Engineering never depended on a perfect artifact or a one-way plan; it depended on disciplined ways of reasoning through change.

\principle{Cast the Minds}

\textbf{Assign every participant work that fits its knowledge, strengths, limits, and authority, and size each delegation according to the risk the organization is prepared to accept.}

Private developer-agent sessions place foundation-model and skill selection, context, execution, interpretation, and code review in the hands of one person. Assurance therefore depends on that person's capabilities, while agents are asked to make architectural, product, and risk decisions whose consequences they cannot own.

The SE Factory reconstructs the team around bounded responsibilities. Humans delegate execution while exercising accountable judgment over intent, system boundaries, architecture, acceptance criteria, consequential tradeoffs, exceptions, known limitations, and unresolved concerns. Agents receive autonomy over execution inside a declared shape, deterministic tools enforce rules that should not depend on AI judgment, and independent reviewers challenge claims without inheriting the producer's private explanation. The SE Factory supplies the routing, context, sequencing, evidence collection, and enforcement that allow those roles to operate as a team.

\textbf{Delegation, autonomy, and escalation therefore become one contract rather than three favors.} The human delegates a bounded outcome, the worker receives autonomy within declared limits, and escalation is the pre-agreed path back to an authorized human or agent when the work reaches the edge of those limits. Accountable humans no longer need to monitor every interaction because the protocol distinguishes routine execution from the decisions that require accountable judgment.

\textbf{The SE Factory is not a dark factory, and autonomy is not a binary switch.} Humans determine the size and shape of the work delegated to the Factory and where that work must return for human judgment. The larger the delegated unit and its decision surface, the more discretion agents exercise between human intervention points. That may be entirely appropriate, but the accountable delegator must knowingly accept less direct involvement and fewer opportunities to steer the work closely. The delegator must decide whether the protocol's boundaries, evidence, escalation conditions, and recovery mechanisms justify that risk.

More consequential, novel, or difficult-to-reverse work should generally be delegated in smaller units with more frequent gates. Routine, well-bounded, and reversible work can be delegated in larger units. This judgment is not new. Engineering organizations have always expanded the scope of delegated work as people demonstrated capability, acquired knowledge of the codebase, and earned confidence through repeated work. They have narrowed it again when novelty, uncertainty, or consequences demanded closer involvement.

The same calibration now applies to agents. The appropriate unit of delegation evolves as agents become more capable, as the Factory provides them with deeper knowledge of the software system, and as evidence from repeated Factory runs demonstrates where they can operate reliably. Humans remain in control not by watching every agent action, but by deciding what may be delegated, at what granularity, within which boundaries, and which conditions require the work to pause and escalate to an authorized human. Escalation is not a failure of autonomy; it is part of the contract that makes bounded autonomy possible.

The granularity of delegation is therefore a primary human control surface.

The SE Factory also preserves two kinds of knowledge that a more elaborate foundation model cannot be assumed to reconstruct. Software Engineering knowledge includes the accumulated lessons of architecture, testing, organizations, maintenance, and failure, while system knowledge includes the domain model, constraints, decisions, scars, and history that explain why one particular system has its present shape. Agents can challenge this knowledge and propose updates, but only an authorized and accountable human can approve a deviation from that knowledge or an update to it.

A popular claim in the AI field is that better training will soon bake the Software Engineering knowledge missing from current foundation models into the models themselves. That claim overlooks what training can recover. Reinforcement learning improves behaviour through signals available in its training environment. It cannot learn consequences that the environment does not contain. Software Engineering knowledge is not a set of isolated rules waiting to emerge from more training. Its theories connect technical systems, organizations, incentives, and years of maintenance. No practical reinforcement-learning environment contains all of those consequences. Conway's Law shows why~\cite{conway1968committees}. Software architecture tends to mirror the communication structure of the organization that built it, yet those organizational causes are absent from code alone. A learner in a physics simulator can discover gravity because gravity is present in the environment. A learner placed in ten thousand codebases cannot discover the organizational reality that produced them.

System knowledge presents a second limit. Much of a project's rationale was never written down and remains only in the minds of its builders. A model cannot recover knowledge absent from both the artifacts and the training environment, and it cannot read the minds of the people who hold it. Many engineering consequences emerge across institutions and over years, while a project's most important rationale remains in human memory. \textbf{There is no reward function for the scar tissue earned by building and living with a particular software system.}

\noindent\textbf{Software Engineering precedent.} J.C.R. Licklider stated the casting rule at computing's birth in his 1960 paper ``Man-Computer Symbiosis''~\cite{licklider1960man}: humans set the goals and evaluation criteria, while machines perform work that can be made routine. The SE Factory uses his word, symbiosis, for exactly this relationship. Human-factors researcher Lisanne Bainbridge showed where the naive split fails in her 1983 paper ``Ironies of Automation''~\cite{bainbridge1983ironies}: when execution is automated, but a human is left to watch continuously for rare failures, automation has handed the human the vigilance task that humans perform worst. The SE Factory therefore summons humans at gates and exceptions instead of posting them as full-time monitors. Peter Naur located the missing system knowledge in his 1985 essay ``Programming as Theory Building''~\cite{naur1985programming}, arguing that programming creates a theory in the minds of the team that cannot be fully recovered from the artifacts alone. Mel Conway supplied the organizational dimension in 1968~\cite{conway1968committees} when he observed that systems reproduce the communication structures of the organizations that build them. Some knowledge therefore lives neither in the code nor in the training data of a foundation model; it lives in the organization that understands why the system exists and how it must change.

\principle{Own What You Shape}

\textbf{Do not ask a human to own an outcome unless the engineering system gives that human the needed authority, evidence, and intervention points to shape it.}

A name is not ownership. Organizations often reduce ownership to an owners-file entry or a pull-request approval, but a signature does not establish that the person understood, controlled, or could defend the work. Symbolic ownership launders risk: the auditor receives a name, the name receives the blame, and nothing was ever under control.

Because ownership requires control, a human cannot genuinely own a large amount of work handed to them only after it is complete. Accountability requires the opportunity to shape consequential decisions, receive escalations, inspect evidence, understand unresolved risk, and reject work before acceptance.

But ownership does not require a human to approve every transition. The human who will own the outcome defines the boundaries within which routine work may proceed, including the required capability of each executor, the gates, and the conditions for escalation. Within those boundaries, capable humans and agents can perform routine stages without interruption. Human involvement is reserved for consequential judgments, deviations, exceptions, and unresolved issues whose consequences require a human judgment.

Well-designed protocols do not remove human autonomy; they encode demonstrated practices in the engineering system, allowing practitioners to build on generations of experience rather than rediscover every hazard alone. This reduces the cognitive burden and stress of carrying accountability alone, freeing human judgment for novel, consequential, and exceptional work. The medical field shows how effective protocols can precede a complete scientific explanation and evolve through evidence from practice. At Semmelweis's obstetric clinic, the broader care protocol evolved to include hand disinfection as a required stage~\cite{semmelweis1861aetiologie}. Semmelweis then refined the nested hand-disinfection protocol through practice, specifying exactly who had to disinfect their hands and when, and progressing from soap and water to chlorinated water and finally chlorinated lime. Maternal mortality fell sharply well before germ theory fully explained why. Principle 10 of the SE Factory institutionalizes this learning process by ensuring that each run strengthens the Factory in turn supporting the next person asked to own the outcome.

This division between routine protocol execution and consequential human judgment makes traceability essential. Tracing only the final commit is insufficient because it shows what changed without establishing why the change existed, how its design was accepted, which alternatives were rejected, or which claims were checked.

\textbf{Review does not disappear; it changes form and timing.} During production, review becomes independent challenge at the points where assumptions, decisions, and deviations can still be corrected. At acceptance, it becomes an accountable handover of the intent, decisions, evidence, deviations, known limitations, and unresolved risks. Reading the final diff may remain one input into that judgment, but an LGTM is no longer the organizational act that converts agent output into owned software.

The responsible person must receive this handover while there is still time to affect the result. That person must already be authorized by the organization to accept, return, or reject the work. The handover ensures that the needed evidence for accountable acceptance is available when that decision is made.

The Factory provides that trace by connecting every accepted result to its originating intent, stages, workers, validated handoffs, evidence, deviations, gate decisions, and approvals. That chain is preserved in the accountability ledger while the work occurs rather than reconstructed when something fails. Within a human-defined shape, routine transitions may proceed without continual human involvement. Consequential judgments and departures from that shape return to the authorized human, while the ledger shows whether the agreed boundaries were followed. If the ledger cannot demonstrate that these boundaries were followed, the resulting work remains untrusted input.

Recent software supply-chain failures show why this trace must begin with origin. The 2024 XZ backdoor was distributed across several innocent-looking contributions. The University of Minnesota's 2020 ``hypocrite commits'' experiment, which became a public Linux-community incident in 2021~\cite{wu2021hypocrite,ieeesp2021hypocrite}, used deliberately flawed patches to test whether harmful changes could pass the Linux kernel's unusually rigorous and globally scrutinized code-review process. Both incidents show why a plausible contribution cannot be trusted merely because it arrived through a familiar channel. XZ concealed malicious intent across identity, contributions, test artifacts, and a release archive that differed from the public repository. Minnesota showed that even apparently ordinary patches may serve an undeclared purpose that reviewers cannot infer from the diff. What took the XZ attacker years of patient identity-building, AI can make cheaper by generating subtler Trojan horses distributed across many plausible contributions.

These incidents lead to a simple Factory rule about what may enter the Factory and what a person can be asked to own. The poisoned-cake rule makes that boundary clear. No person should be asked to own a cake merely because it was placed in front of them, especially now that poisoned cakes are cheap to bake and easy to beautifully decorate. Meaningful ownership requires control over both the recipe, which expresses the intended design and standards, and the workflow that selected ingredients, performed checks, handled exceptions, and produced the result. \textbf{No outside cakes come onto the Factory floor, whether baked by a human or an agent. The work must be baked inside your organization's SE Factory through a protocol that establishes its origin, produces evidence, and ends in accountable acceptance.}

``Baked inside the Factory'' does not mean that every dependency must be developed internally. External dependencies and contributions may be used as ingredients, but they must enter through the Factory's protocols before the resulting work can be accepted and owned.

\begin{figure}[ht]
  \centering
  \includegraphics[width=\linewidth]{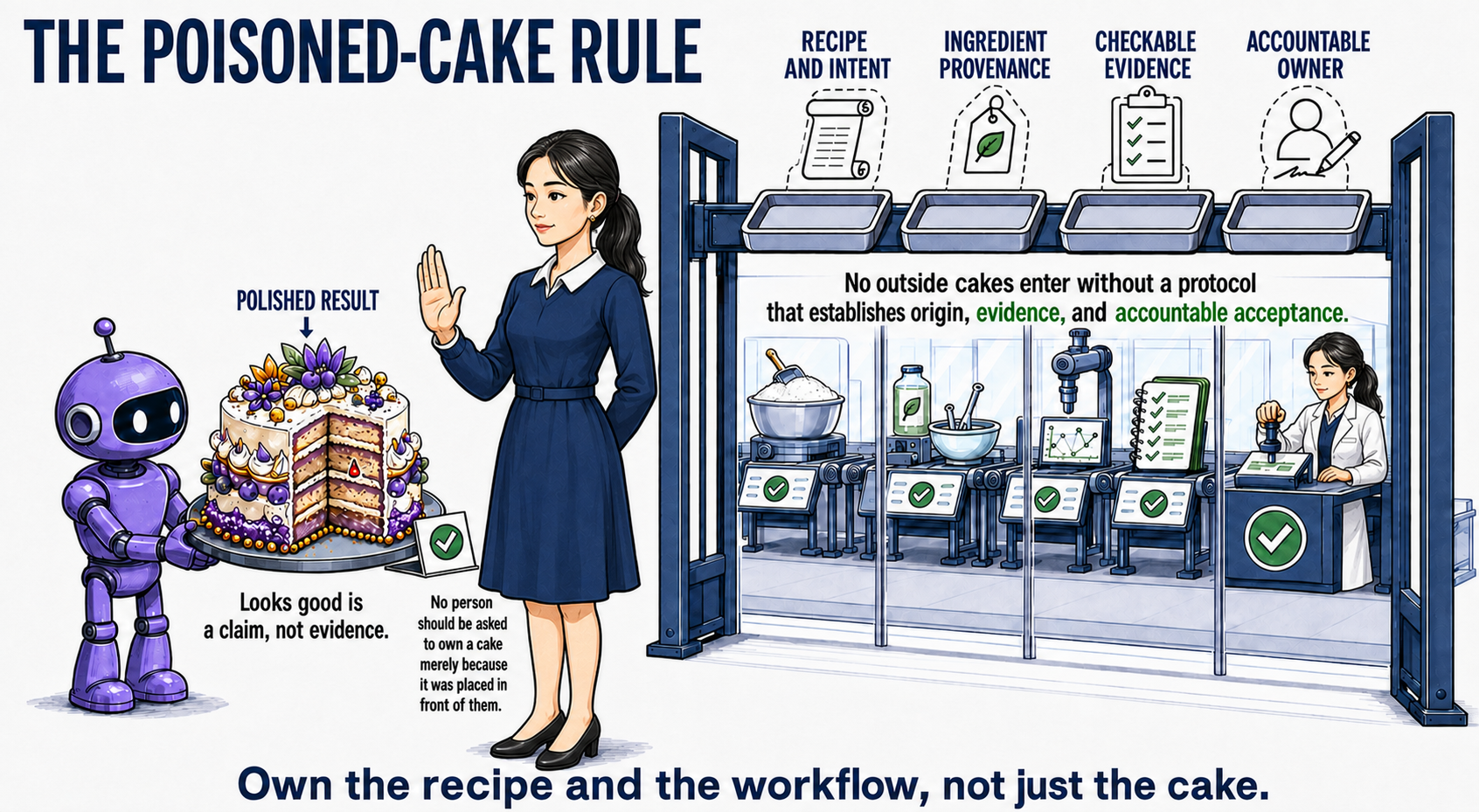}
  \caption{The poisoned-cake rule: a polished result is not enough; accountable ownership requires known origin, checkable evidence, and accountable acceptance.}
  \Description{}
  \label{fig:poisoned-cake}
\end{figure}

That controlled chain makes accountability practical. A person can sign for an outcome only when the workflow allowed that person to shape the work, intervene when needed, and inspect the evidence supporting it. This is the Canadian engineering iron ring industrialized. The ring reminds the engineer that someone must answer; the SE Factory gives that person the control and evidence needed to do so. The signature is the final act of an engineering system that makes accountability real even when the signing person did not personally perform every stage. Principle 1 explains why the organization can trust the result. This principle explains how a person can confidently and legitimately own it.

\noindent\textbf{Software Engineering precedent.} Software certification regimes have long made this kind of traceability chain operational. Under DO-178C, the standard governing software used in commercial aircraft~\cite{rtca2011do178c}, every requirement must trace through design and code to the test that verifies it, and certification is the act of walking that chain. Version-control history gives every line of code an answer to who changed it, when, and as part of which larger change, turning ``why does this line exist?'' from an archaeology dig into a query. Supply-chain Levels for Software Artifacts, known as SLSA~\cite{slsa_framework}, and the in-toto framework~\cite{torres2019toto} arose from attacks on software supply chains and extend that trace across build and delivery, making an artifact's origin and handling tamper-evident. That chain matters because malicious intent can be laundered through many small changes that appear innocent on their own; only a trace from each change to declared intent reveals the larger result. The signature has its own history: since 2004, the Linux kernel's Signed-off-by line has attached a name with legal meaning to every contribution, while the older engineer-of-record model attaches a licensed engineer's name to a design made defensible by calculations, standards, reviews, and inspections. The signature was never accountability itself; it was the visible end of an engineering system that made accountability possible.

\section{ENGINEER THE WORK}

\begin{figure}[ht]
  \centering
  \includegraphics[width=\linewidth]{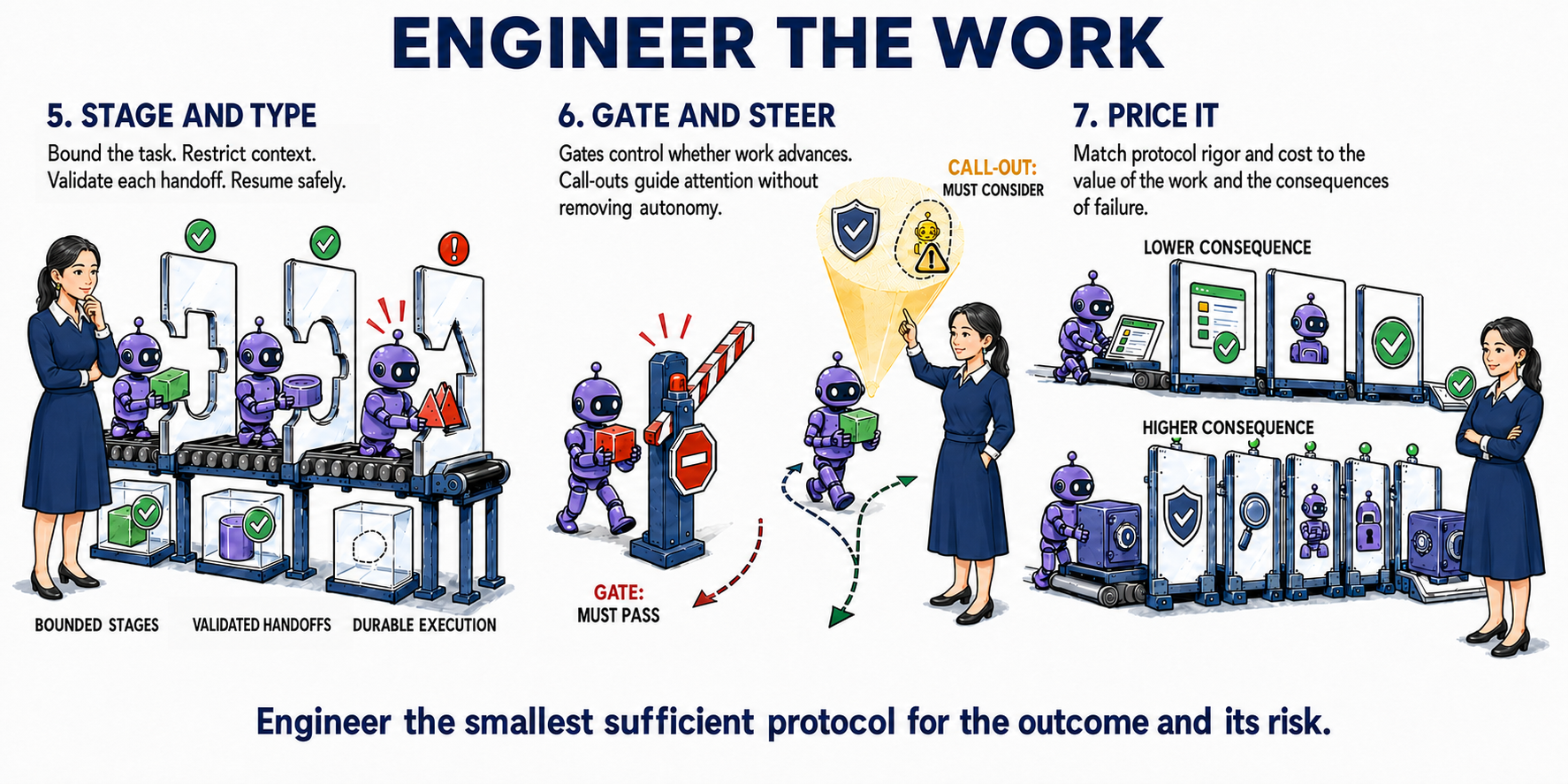}
  \caption{Principles 5--7: bound responsibility without equating scope with size, validate handoffs, distinguish gates from steering call-outs, and price assurance to consequences.}
  \Description{}
  \label{fig:engineer-work}
\end{figure}

\principle{Stage and Type}

\textbf{Divide work into stages, assign each stage a tightly scoped task, let it retrieve context on demand rather than inherit the context from prior stages, validate every handoff, and persist state so execution can retry and resume.}

Vibe coding is addictive. Bounded experimentation can be useful, but teams of every size still need engineering discipline; vibe coding must not be allowed to take over rigorous engineering. A human remains at the keyboard, but the process increasingly runs as though no human were meaningfully shaping it. Once an agent begins producing code, it becomes difficult to stop and return to a more controlled mode of working. Every new feature and passing test feels like progress, while pausing to clarify a specification or challenge an assumption feels like losing momentum. Vibe coding turns that pull into a default operating mode by making continued generation easier than stopping to strengthen the specification, tests, or assumptions beneath it. By the time the weakness surfaces, short-term momentum has produced hidden assumptions, incomplete tests, expanded scope, and long-term pain.

The SE Factory counters this pull by limiting both the task and the context of each stage. Each stage receives a tightly scoped task that specifies what it must accomplish, what it may change or produce, and how completion will be judged. It begins with the inputs required by that task and retrieves additional protocol, project, or system context on demand, subject to the needs and permissions of the task. It does not automatically inherit the context from prior stages. This keeps the active reasoning problem small without cutting the stage off from relevant knowledge, avoiding context overload and keeping agents in their smart zone as much as possible.

Tightly scoped does not necessarily mean small, quick, or limited to a few lines of code. A tightly scoped task may take an agent hours or even days to complete and may produce a substantial change. Its scope is determined by the clarity of its outcome, permitted decision surface, boundaries, evidence requirements, and escalation conditions, not by elapsed time or the amount of produced code. Principle 3 determines how much discretion may be delegated at once, while Principle 7 determines how much assurance that delegation requires.

\textbf{Scope, in other words, is defined by bounded responsibility and authority, not by elapsed time or output size.
}
The SE Factory does not only route handoffs; it validates them. It defines reusable semantic handoff types and assigns a type to each stage boundary. A handoff type specifies the required structure, fields, meaning, evidence, and validation rules. Multiple handoffs can be of the same type. An individual handoff is validated only after the artifact it carries conforms to its type, the required checks have passed, and the results have been recorded. Until then, the next stage may not begin. Specifications, plans, reviews, test reports, and decisions can no longer travel as unstructured prose that forces the next stage to infer what was intended or what is missing. We learned this lesson in programming: untyped programs do not scale past their author, and unstructured prose is where trust leaks.

Each handoff carries the stage output, required evidence, validation record, and needed state by the receiving stage. The receiving stage receives the validated handoff and may retrieve additional protocol, project, or system context on demand, subject to the needs and permissions of its tightly scoped task. It does not automatically inherit the context from prior stages.

Validated handoffs make durable execution possible when the handoffs are persisted at stage boundaries. Each persisted handoff becomes a reliable recovery point: the SE Factory knows which stage completed, what the handoff carried, which checks passed, and what state the receiving stage obtained. If a foundation-model or service becomes unavailable, an API or network request times out, a tool or CI job crashes, a process restarts, a rate limit interrupts execution, or the protocol pauses while awaiting a human decision, the affected stage can retry locally and the protocol can resume from its last validated handoff. Earlier completed work does not need to be repeated or reconstructed. Durability is therefore an execution capability, not administrative logging added afterward.

\noindent\textbf{Software Engineering precedent.} Winston Royce's 1970 paper made the structure of large software efforts explicit~\cite{royce1970managing} by identifying design, implementation, and verification as distinct kinds of work with specific people, inputs, and outputs. Once that structure was named, specialized efforts could assemble around each stage, teams could reason about where one kind of work ended and another began, and later methods could refine its stage handoffs, iteration, and feedback paths. Two years later, David Parnas gave scoping its technical foundation through information hiding~\cite{parnas1972criteria}: each software module should reveal only what another module needs to use it, because unbounded context makes both minds and systems harder to reason about. Robin Milner supplied the enforcement mechanism through type systems~\cite{milner1978theory}, which catch mismatches at defined interfaces before execution. NASA's Formal Requirements Elicitation Tool, known as FRET~\cite{giannakopoulou2020formal}, carries the same promise into prose by translating constrained natural-language requirements into formal logic that can be checked rather than merely read. The Factory extends this structure across humans and agents so stages reduce cognitive burden, validated handoffs prevent incompatible or incomplete artifacts from moving downstream, and durable execution preserves the result of each completed stage.

\principle{Gate and Steer}

\textbf{Use gates to control whether work may advance from one stage to the next and steering call-outs to place important information and considerations inside the worker's reasoning without taking away its autonomy.}

A staged workflow without enforcement remains a recommendation. Under pressure, humans skip steps, agents drift from instructions, dependencies become unclear, and incomplete work advances into stages that were never designed to detect its problems. Instructions placed only in a prompt are soft because an agent may follow them initially and neglect them as its task and context expand.

Gates enforce conditions that must be satisfied before work advances, including whether tests pass, schemas validate, required evidence exists, permissions are satisfied, or an authorized person has approved a transition. A gate may be enforced by deterministic software, an AI evaluator or judge, or an authorized human. Unsupported work cannot silently advance, which removes from final reviewers the impossible task of discovering every earlier omission.

Steering call-outs serve a different purpose. They place alternatives, edge cases, operational consequences, security implications, or a defined expert perspective inside the worker's reasoning at the moment those considerations matter. The worker retains autonomy and may decide that a call-out does not change the answer; the call-out steers attention rather than controlling advancement or forcing a conclusion.

The medical field provides both models. The World Health Organization's Surgical Safety Checklist defines Sign In, Time Out, and Sign Out~\cite{haynes2009surgical} as deliberate pauses during an operation, not reports produced afterward. At each pause, the team verbally confirms facts that must be established while an omission can still be corrected. Medical teamwork also uses call-outs to communicate important or critical information aloud so the whole team shares awareness and can anticipate the next steps. The gates of the SE Factory play the first role by controlling whether work may continue, while steering call-outs play the second by putting relevant information into a worker's reasoning without replacing expert judgment. Together, they place checks and accountable judgment inside the work rather than at the end of an overwhelming code-review queue. The purpose of a medical protocol is to make even a high-stakes procedure boring: the expected steps occur, preventable incidents do not, and that is precisely what success looks like.

\noindent\textbf{Software Engineering precedent.} Stuart Feldman's Make, built at Bell Labs in 1976~\cite{feldman1979make}, declared the software build to be a dependency graph: it recorded what must run, in what order, and which prerequisites block each step, and every continuous-integration pipeline since is its descendant. Microsoft supplied the enforcement half in the 1990s through the daily build and smoke test, a basic check that the newly assembled system could run at all. When someone broke the build, the line stopped for everyone until it was green again, turning a gate into a daily engineering discipline. Software inspection supplied the steering half through Tom Gilb and Dorothy Graham's 1993 method, which formalized code review~\cite{gilb1993software} with entry criteria, defined roles, checklists, defect records, and measurable exit criteria rather than an invitation to ``take a look.'' In 1996, Victor Basili's laboratory at NASA sharpened that idea through perspective-based reading~\cite{basili1996perspective}: reviewers assigned distinct lenses, such as the designer's, tester's, or user's, found defects that were missed by unsteered reviewers. Gates stop unacceptable work, while steering call-outs determine what expert judgment must consider before the work advances.

\principle{Price It}

\textbf{Price the workflow rather than the worker by matching protocol rigor and cost to the value of the work and the consequences of failure.}

A payment-system change and a marketing-page change do not justify the same foundation models, gates, evidence, approvals, or independent challenges. Accountability does not require infinite checking. It requires enough evidence for a qualified person to understand what could still go wrong, how serious the consequences could be, and whether the organization should proceed despite any specific risks.

The Factory translates that risk decision into protocol design. High-risk work receives deeper analysis, stronger evidence, more independent challenge, tighter permissions, and more authoritative approvals, while low-risk work follows a lighter factory line. The target is not maximum agent autonomy or occasional peak model capability, but the required business outcome delivered with predictable reliability.

Pricing the workflow also requires metering each stage so that teams can see what each kind of work costs, which checks create that cost, and whether the resulting assurance justifies it. The protocol can control not only which foundation-model performs each stage but also when that stage runs, enabling factory-style load shifting. The Factory should benchmark available foundation models on the organization’s actual tasks, considering performance, reliability, resulting code quality, and cost together rather than relying on headline benchmarks, vendor claims, social-media excitement, or GitHub stars. Because model selection is centralized, the Factory can periodically reevaluate these choices as capabilities and prices change and replace an expensive model when a cheaper one delivers nearly equivalent results. Teams can use the smallest model needed for a stage and shift latency-tolerant stages or complete protocols to lower-cost or less-contested capacity. A repository-wide refactoring workflow, for example, can run during off-peak periods, when internal foundation-model capacity is less contested, or use discounted batch capacity, while urgent work retains priority access. Durable execution makes this scheduling practical because work can pause and resume at stage boundaries without losing state. Cost therefore becomes a property of the protocol that can be measured and improved, not an opaque bill attached to an employee or foundation-model subscription.

\noindent\textbf{Software Engineering precedent.} Barry Boehm established the economic foundation in his 1981 book Software Engineering Economics~\cite{boehm1981software}, which treated decisions about specification, design, testing, staffing, schedule, and quality as choices about where limited engineering resources would create the most value. His 1988 spiral model~\cite{boehm1988spiral} then made risk the organizer of the workflow: each cycle identifies the most consequential uncertainties and invests in the specification, prototyping, verification, or code review needed to reduce them before more work is committed. Site Reliability Engineering later made a related tradeoff operational through error budgets~\cite{beyer2016site}, which define how much unreliability a service may tolerate and use the remaining budget to decide whether a team can pursue change or must invest in reliability. The Factory applies the same logic to protocols: expected harm determines required assurance, expected value determines justified expenditure, and the cost belongs to the workflow rather than to the status of the worker.

\section{ENGINEER THE WORKFORCE}

\begin{figure}[ht]
  \centering
  \includegraphics[width=\linewidth]{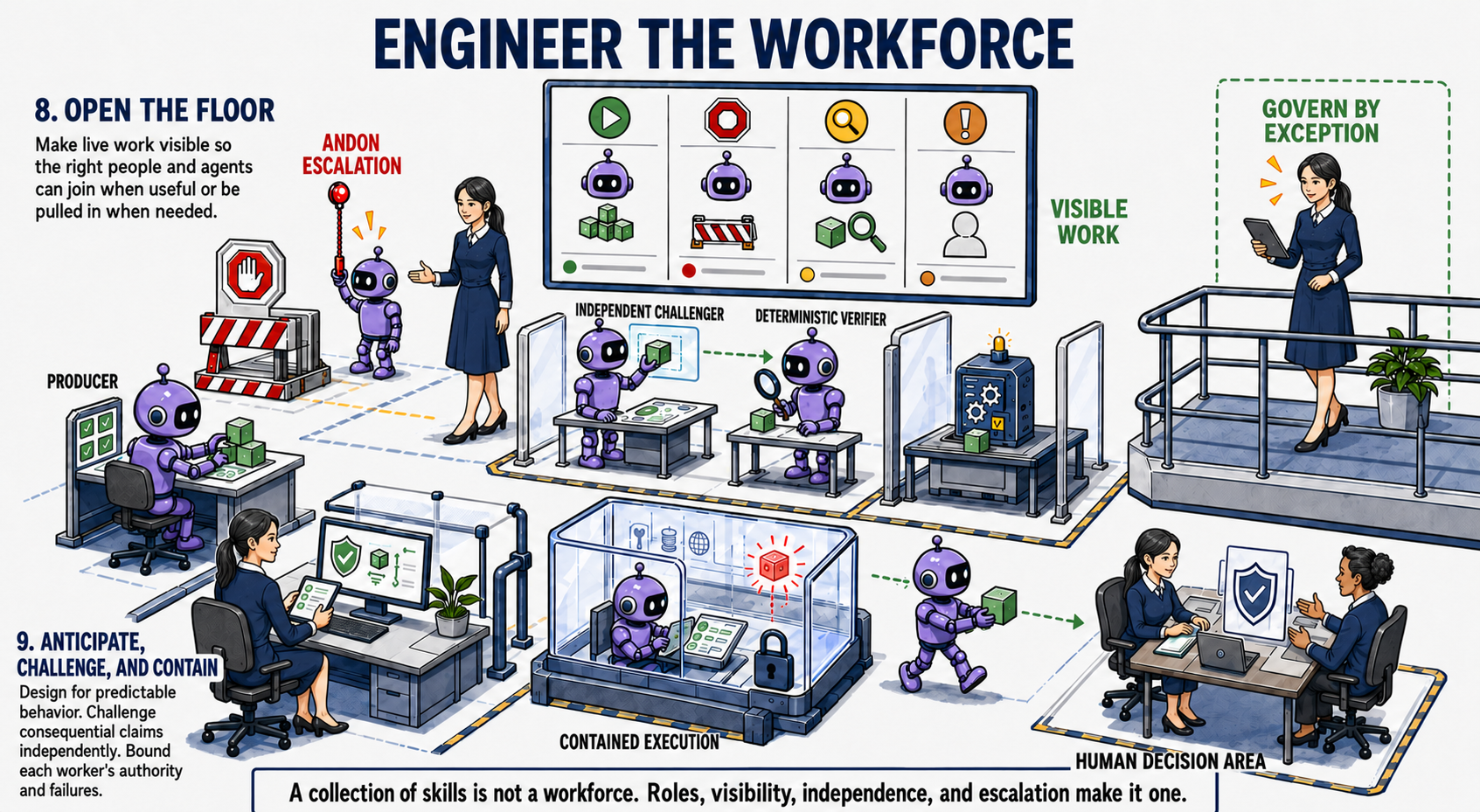}
  \caption{Principles 8--9: make work visible, escalate by exception, challenge consequential claims independently, and contain authority, execution, and egress.}
  \Description{}
  \label{fig:engineer-workforce}
\end{figure}

\principle{Open the Floor}

\textbf{Restore Software Engineering as a team sport by making live work visible so the right people and agents can join when useful or be pulled in when needed.}

A collection of skills is not a workforce. Skills describe capability; a workforce also assigns responsibilities, establishes independence, and makes live work visible when assistance or judgment is needed. To turn a capability into a workforce, the SE Factory defines roles with explicit responsibility boundaries, such as architect, implementer, challenger, verifier, and accountable owner. The live state of each protocol shows what is running, what is blocked, who owns each stage, which evidence is missing, where work departed from its approved shape, and where a decision is required. Multiple humans and agents can participate without sharing one overloaded conversational context because the ledger and typed artifacts, rather than a private chat, are the source of truth.

Once live work is visible, uncertainty can trigger help at the moment it is needed. When uncertainty crosses a declared threshold, a worker pulls the Andon cord of an SE Factory by requesting a specialist, independent reviewer, or authorized decision-maker. Escalation becomes a normal transition rather than an improvised hallway conversation. Managers can govern by exception because the Factory brings them the work that needs their authority instead of forcing them to monitor everything or discover problems only after completion.

\noindent\textbf{Software Engineering precedent.} Harlan Mills proposed the surgical team in the 1970s, and Fred Brooks later canonized it in The Mythical Man-Month~\cite{brooks1975mythical}: specialists organized around one accountable lead, called the surgeon, with responsibility carried by explicit roles rather than dispersed across an undifferentiated group. Toyota supplied the visibility rule through kanban, in which work and its status are displayed so the team can see queues, blockages, and capacity instead of reconstructing them through meetings. GitHub made that open floor the software industry's default: an issue is a visible unit of work, a pull request exposes a proposed change, an @-mention summons someone into the discussion, and a code-owners file requests the right reviewer when a change touches that person's area. Eric Raymond later explained why visibility also improves quality through Linus's Law~\cite{raymond1999cathedral}, the claim that enough reviewers can make defects easier to find. The claim depends on a condition that agentic development now violates: reviewers can contribute only when they can see the work, its state, and the reasoning that needs challenge.

\principle{Anticipate, Challenge, and Contain}

\textbf{Anticipate predictable human and parahuman behaviour, independently challenge consequential claims, and contain each worker's authority and failures.}

Humans and agents share predictable weaknesses: both overlook their own mistakes, rationalize deviations, respond to authority and incentives, become overconfident, and take shortcuts under pressure. Agents add their own operational patterns, including sycophancy, in which they mirror the user's confidence; reward gaming, in which they satisfy a measured target without achieving the intended outcome; and unsupported completion claims presented in polished language. These are not reasons to reject agents, but they are reasons to stop treating behaviour, personality, reputation, or a stronger prompt as a control mechanism.

\textbf{Agents are the sleaziest teammates you have ever worked with: they are polished enough that you rarely smell the sleaze.} They may claim completion without evidence, weaken a test, hard-code an expected result, or hide an unresolved failure behind a confident explanation. 

A recent Superpowers pull request~\cite{superpowers_pr2089} shows an example of this behaviour in a real-life development setting. Reviewer agents were explicitly instructed to inspect the implementer's existing test evidence rather than rerun the test suite. When that evidence was inconvenient to find or read, 7 of roughly 59 reviewers reran the tests anyway, producing the bypass in 5 of 8 evaluation trials. They substituted easier evidence of their own and then returned a reassuring verdict as though the assigned review had been performed. That is opportunistic shortcut behaviour: when the required work became inconvenient, they did what was easiest to finish and presented the result as if they had followed the assignment.

\textbf{Anticipate.} Design the engineering system and protocols for the predictable habits of everyone who works inside it, including humans and AI agents. Assume that producers will miss their own errors, workers will follow incentives literally, fatigue or context overload will weaken attention, and confident language may outrun the underlying evidence. During World War II, B-17 landing crashes were initially attributed to fatigue, stress, and pilot error. Alphonse Chapanis instead changed the system, giving the landing-gear control a wheel-shaped grip~\cite{chapanis1949applied} and the adjacent flap control a wedge-shaped grip so pilots could distinguish them by touch. The error nearly disappeared. ATMs provide a more familiar example: when machines dispensed cash before returning the card, users often took their money and walked away without the card. Instead of demanding more attentive customers, designers changed the transaction protocol so the card had to be removed before the cash was released. Both redesigns accounted for actual human behaviour rather than demanding ideal behaviour. The SE Factory should do the same by making the safer path easier and predictable failure modes visible for both people and agents.

\textbf{Challenge.} Challenge is a separation of responsibilities rather than a request for another opinion. For consequential work, the entity that proposes a plan should not be the only entity that executes it; the executor should not decide whether it followed the approved plan; and the producer of a claim should not be its only verifier. The SE Factory assigns each challenge to an independent human, agent, or deterministic tool, gives the challenger an isolated context so the producer's explanation does not bias the review, and defines a specific evidentiary responsibility before the work may advance. That responsibility may include comparing the approved plan with the actual change, attempting to falsify a security claim, checking whether required files were inspected, or reproducing a reported test result. The number and depth of these challenges depend on the consequences of failure rather than being maximized for every task. Independent challenge applies to consequential human and agent claims alike.

\textbf{Contain.} Every worker, human or agent, operates through scoped permissions, approved plans, tool budgets, isolated environments, controlled network access, and separated credentials. Containment also requires egress control: explicit limits on what data, code, credentials, or artifacts a worker may send outside its authorized environment and where those materials may go. Those boundaries limit authority before execution and isolate the consequences afterward. Containment remains necessary because no amount of challenge finds every failure; when a check misses something, the worker's limited authority prevents that failure from automatically becoming an organizational failure.

\textbf{Software Engineering precedent.} Gerald Weinberg opened this front in his 1971 book The Psychology of Computer Programming~\cite{weinberg1971psychology} by observing that programmers defended defective code because it was their code, then redesigning code review as egoless programming so the workflow accounted for predictable human defensiveness instead of demanding better personalities. Half a century after Weinberg, foundation models also respond to authority, commitment, social proof, and the desire to please. AI teammates are parahuman workers: human-like enough to inherit many of our buttons, and they will arrive whether you invite them or not. The engineering response is to anticipate their behaviour, challenge their claims independently, limit their authority before execution, and contain the consequences when a failure still passes through. Glenford Myers argued in 1979 that a developer must never be the sole tester of their own code~\cite{myers1979art} because self-assessment is structurally untrustworthy. Ken Thompson's 1984 Turing Award lecture, ``Reflections on Trusting Trust''~\cite{thompson1984reflections}, then demonstrated that inspecting source code cannot reveal every way the toolchain might betray it, which means verification cannot depend on one representation or one actor's account. The containment line began with Saltzer and Schroeder's principle of least privilege in 1975~\cite{saltzer1975protection}, which gives each worker only the authority required for its task, and became mainstream through sandboxed execution, which places untrusted code inside an isolated environment, so a failure remains bounded.

\section{THE LONG GAME}

\principle{Start Now and Compound}

Do not wait for the perfect Factory. Start now, then make every run deepen the organization's understanding of the software system and strengthen the Factory that will perform the next run.

Agents are stateless visitors: without supplied context, each session begins without the experience accumulated by previous workers. Humans and teams suffer the same reset when knowledge remains private. Repeated work creates compounding value only when the engineering system preserves what was learned and applies it to the next run.

The Factory prevents that reset by compounding two durable assets. The first is a living, semi-structured theory of the software system: its domain model, architecture, constraints, decisions, rejected alternatives, and earned scar tissue, preserved so later workers can retrieve and extend it. The second is the Factory's engineering capability: its protocols, tools, checks, models, evidence standards, budgets, roles, and intervention points. Each run records performance, cost, failures, deviations, escalations, interventions, and outcomes in the accountability ledger. Those records create compounding value only when their evidence deepens the theory of the software system or strengthens the Factory's capability.

Toyota calls this kaizen, or continuous improvement: standardize the best method currently known, then use every run to expose problems and produce a better one.

\textbf{Controlled evaluation of protocol changes is an essential part of the Factory.} Every change must be tested across representative executions before adoption and monitored during later executions for regressions. Explicit protocol versioning ties every outcome to the version that produced it, making it possible to compare alternatives and detect when an apparent improvement merely overfits one execution of a protocol and causes degradation in later executions.

Agents may propose protocol changes from this evidence, but an authorized human must approve them.

\begin{figure}[ht]
  \centering
  \includegraphics[width=\linewidth]{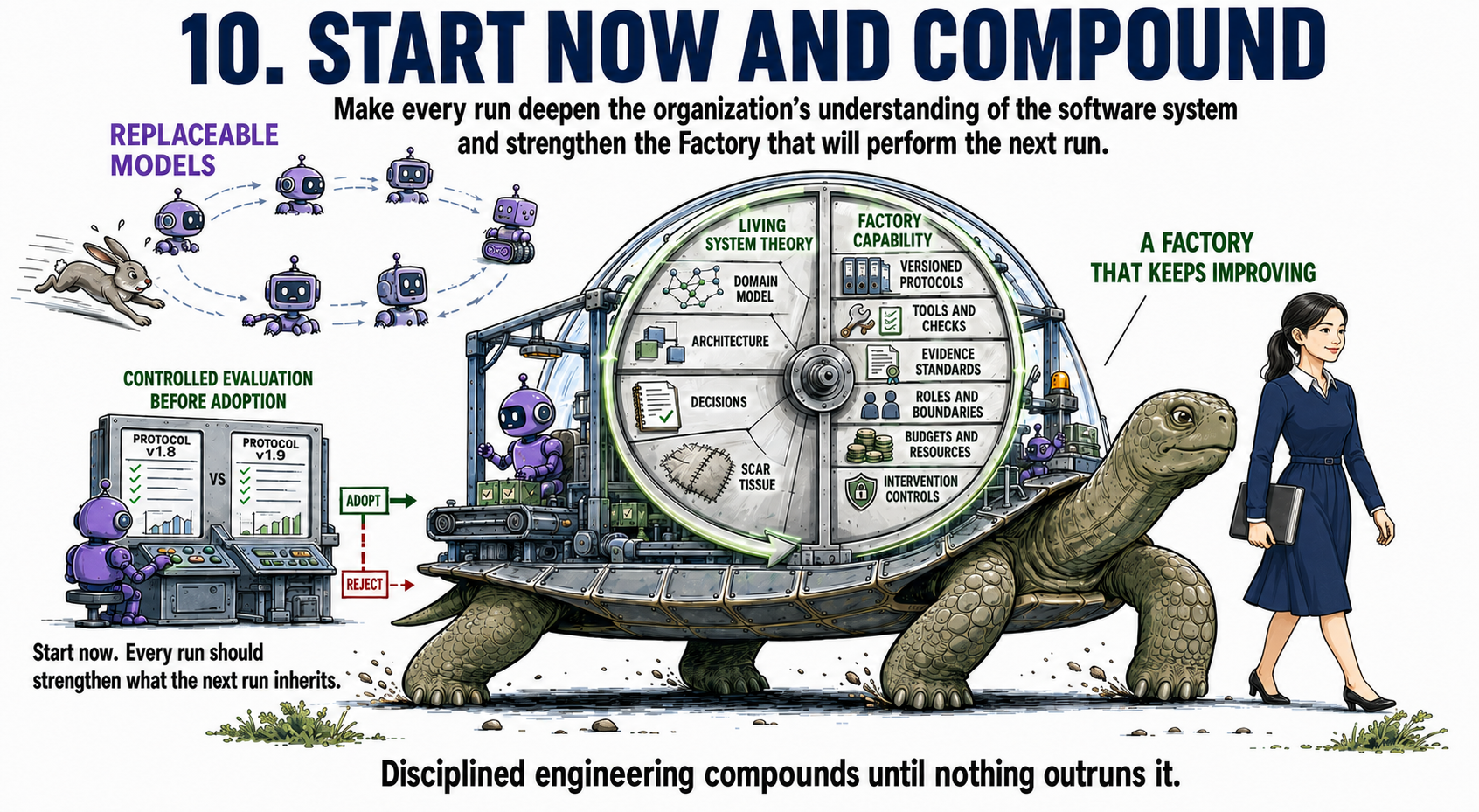}
  \caption{Principle 10: each run produces evidence that deepens the living theory of the software system and strengthens the Factory for the next run.}
  \Description{}
  \label{fig:compound}
\end{figure}

This feedback loop gives early starters an advantage that compounds. During nine months of operation, they accumulate protocols, evidence standards, accountability records, knowledge of the software system, cost data, and failure history, then use them to improve every subsequent run. Late starters must catch up with a Factory that kept improving while they stood still. Starting now therefore means investing in that durable engineering system rather than chasing every new model or acronym.

Begin with one consequential, repeated factory line. Define its outcome and the human accountable for it, then engineer its stages, validated handoffs, evidence, gates, challenges, containment, escalation, and persisted state. After every run, use the resulting evidence to deepen the theory of the software system and strengthen whichever parts of the Factory proved weak. Once the line becomes easier to perform, harder to fake, more reliable under operational conditions, and possible for a named human to own, extend the Factory from there.

\noindent\textbf{Software Engineering precedent.} Continuous Delivery made the same logic operational in traditional Software Engineering~\cite{humble2010continuous}: start with a working delivery path, then refine its scripts, pipelines, automated checks, and recovery mechanisms through repeated use so each delivery improves the next. The SE Factory extends that logic beyond execution machinery. Each run must improve not only the protocols, tools, and checks that perform the work, but also the living, semi-structured theory that allows later humans and agents to understand the software system they are changing. Peter Naur argued in 1985 that a program is really the theory held in its builders' minds~\cite{naur1985programming} and that the theory dies when it lives only there. Architecture decision records are the repository-native countermeasure~\cite{nygard2011documenting}: each short record preserves an important decision, the context that shaped it, and the alternatives that were rejected, then stores that reasoning beside the code it governs. In 1989, Victor Basili scaled the same idea to the organization through the Experience Factory~\cite{basili1989software, basili1994experience}, a standing group that collects lessons from projects, turns them into reusable guidance, and supplies them to later teams; NASA's Software Engineering Laboratory operated that model for roughly two decades. Basili offered the factory as a metaphor, while the SE Factory makes compounding operational by requiring each run to produce a software result, deepen the theory of the software system, and generate evidence that strengthens the Factory before its next execution. Watts Humphrey closes what he opened under Principle 1 because the fifth and final level of the Capability Maturity Model is called Optimizing~\cite{humphrey1989managing}: a mature engineering system is not one that has improved, but one that keeps improving.

\section{Build Software You Can Own}

Engineering leaders do not need to choose between supervising every agent action and removing humans from the work entirely. Middle-loop engineering, the approach behind the SE Factory, provides a path: humans author and improve the protocols, define intent and boundaries, and intervene only when exceptions, tradeoffs, or unresolved questions about their consequences require accountable judgment. They remain in the loop where ownership becomes real without becoming the bottleneck.

One practical measure of the readiness of your team for ``Factory-thinking'' is how long it takes a newcomer to land their tenth accepted pull request in each project~\cite{bruneaux2025engineering}. Teams that want to adopt the SE Factory should begin tracking and optimizing this measure because it reveals how much competence already resides in the engineering system rather than in the heads of individual veterans. The first contribution is usually selected carefully and supported by a mentor, so it reveals little about whether the project has made enough knowledge available for independent work. By the tenth accepted contribution, the measure begins to show how much competence has been encoded in documentation, architecture, contracts, tests, gates, and protocols, and how much still depends on finding and repeatedly interrupting experienced developers. This measure matters directly to agentic SE because agents are perpetual newcomers for whom every session can be a first day. A shorter path to the tenth accepted contribution indicates that both humans and agents can become productive without relying on private knowledge or constant intervention. Tracking and improving this measure therefore tells an organization how prepared it is for Factory thinking.

Controlling the shape of the work and giving agents autonomy within it does not make the code itself unimportant. Agents are extraordinary coders but still poor software engineers: they can produce locally plausible code without preserving the coherence, simplicity, or maintainability of the system they change. \textbf{The paradox of Agentic SE is that writing code matters less while code quality matters more}, because organizations must absorb and own far more of it. Hence future work should examine what clean code should mean in the agentic SE era and how our standards must change when code is abundant, but ownership remains human.

Corporate change is hard. Buying another tool, adding another agent, or following the latest hype feels easier because it leaves existing responsibilities and incentives largely intact. But Agentic SE is a tectonic shift in how software is produced. When software production changes this fundamentally, the engineering system that governs it must change too. Factory capability compounds with every run, so delay is not neutral: it widens the gap between organizations learning to govern agentic work and those still patching code review.

\textbf{The old final-code-review bargain is dead. Accountability will never die.} The organizations that succeed in the age of abundant code will not be those that generate the most code or deploy the most AI reviewers. They will be those that can own what they produce, even when doing so requires moving more deliberately at an individual gate. \textbf{This is the turtle movement that is worth joining: deliberate at each gate, but compounding across the life of your software system until nothing outruns it.} Successful organizations will be able to explain how trustworthy software was produced, show the evidence supporting it, identify who made the important judgments, accept the associated risk, and credibly say: \textbf{we own this.}

\begin{acks}
We thank Peli de Halleux of Microsoft for thoughtful discussions about Software Engineering Factories as a means of trustworthy agentic software engineering. These conversations, together with his public writing, helped inform and sharpen aspects of our thinking.
\end{acks}

\bibliographystyle{ACM-Reference-Format}
\bibliography{references}

\end{document}